# Dynamics of the spontaneous emission factor in multiple quantum well nanowire lasers

Parya Reyhanian
School of Electrical and Data Engineering, University of Technology Sydney, Ultimo NSW 2007 Australia
Christopher G. Poulton
School of Mathematical and Physical Sciences, University of Technology Sydney, Ultimo NSW 2007 Australia
Arti Agrawal
School of Electrical and Data Engineering, University of Technology Sydney, Ultimo NSW 2007 Australia
arti.agrawal@uts.edu.au
Charlene J. Lobo
School of Mathematical and Physical Sciences, University of Technology Sydney, Ultimo NSW 2007 Australia
charlene.lobo@uts.edu.au

***Abstract*—The spontaneous emission factor - often known as the $\beta$ factor - is an important quantity in the description of quantum well lasers, influencing both the threshold power as well as the general shape of the light in-light out (L-L) curve. Past work on modelling multiple quantum well (MQW) nanowire laser devices has typically assumed that the $\beta$ factor is a constant parameter that can either be estimated or fit in a post-hoc manner. However, the $\beta$ factor can be derived from the transitions between valence and conduction bands in semiconductor quantum wells, together with knowledge of the cavity modes. Here we investigate the dynamic nature of the $\beta$ factor for MQW nanowire lasers, and show how it can be computed. We also examine the dependence of the spontaneous emission rate and spontaneous emission factor $\beta$ on the charge carrier density, quantum well thickness, and composition, and discuss the impact on laser threshold and operation.**

## I. Introduction

Multiple quantum well (MQW) nanowire lasers stand out for their compact footprint, energy-efficient operation, and strong potential for on-chip photonic integration [1]. Two defining features, namely the step-like density of states and the strong optical confinement within radial heterostructures, enable sharp gain spectra and efficient carrier recombination [2], [3]. These characteristics not only reduce lasing thresholds but also lead to enhanced spontaneous emission rates compared to conventional bulk semiconductor lasers, making MQW nanowire lasers exceptionally well-suited to next-generation optoelectronic and quantum photonic devices [4].

In MQW nanowire lasers, the spontaneous emission dynamics are strongly influenced by confinement of charge carriers in the quantum wells, which both modifies the photonic environment and the photonic density of states (PDOS). Combined with the improved spatial overlap between electrons and holes, this leads to a significantly higher spontaneous emission rate compared to conventional bulk semiconductor lasers [4]–[6]. The enhancement in the spontaneous emission can be quantified by evaluating the $\beta$ factor, which was defined by Yamamoto *et al.* as the ratio of the spontaneous emission rate into the lasing mode divided by the total spontaneous emission rate [7].

The general expression for $\beta$ is given as the ratio between spontaneous emission into the lasing mode to the total spontaneous emission into all modes, including the continuum of free-space and leaky modes surrounding the cavity:

$$\beta = \frac{R_{\mathrm{sp,cav}}}{(R_{\mathrm{sp,cav}} + R_{\mathrm{spon}}^{\mathrm{cont}})}, \tag{1}$$

Here, $R_{\mathrm{sp,cav}}$ is defined as the spontaneous emission rate into the cavity mode, and $R_{\mathrm{spon}}^{\mathrm{cont}}$ is the spontaneous emission rate into the continuum of all other non-resonant modes.

Calculating $\beta$ properly involves complex evaluations of the dynamics of the optical processes within the laser structure [8]. Therefore, most prior work has avoided this by treating $\beta$ as a constant [9]–[11]. Yamamoto evaluated $\beta$ in microcavity lasers as follows:

$$\beta = \frac{\lambda^4}{4\pi^2 V \Delta\lambda \epsilon^{(3/2)}}, \tag{2}$$

where $\lambda$ is the emission wavelength, $V$ is the volume of the cavity, $\Delta\lambda$ is the spontaneous emission linewidth, and $\epsilon$ is the permittivity [7]. Other researchers calculated the radiative decay rate in nanowire lasers with constant terms $F\beta$ for the lasing mode and $(1-\beta)$ for background radiation [9], [12].

Gregersen [13] developed a self-consistent nanolaser framework that explicitly treats both optical and electronic densities of states. To incorporate the dependence of $\beta$ on charge carrier density, Romeira *et al.* [8] computed $\beta$ for InP micro- and nanopillar and metallo-dielectric cavity geometries with InGaAs bulk gain medium from microscopic spontaneous emission rates. These previous authors noted that $\beta$ would depend on the charge carrier density, but assumed a fixed rate of spontaneous emission into the continuum of free space modes.

Because both the cavity-enhanced and background radiative rates depend on carrier density through band-filling and Fermi–Dirac statistics, $\beta$ becomes a dynamic, carrier-density-dependent quantity set by the balance between Purcell-enhanced emission into the lasing mode and leakage into non-lasing channels. In this definition, a reduction in effective mode volume [14] ($V_{\text{eff}} = \frac{\int \epsilon(r)|E(r)|^2 dr}{max(\epsilon(r)|E(r)|^2)}$) strongly increases $\beta$, so that for the smallest nanopillar cavity, $\beta$ approaches unity and yields an almost thresholdless light–current characteristic, whereas larger cavities retain $\beta \ll 1$ [8]. Such high $\beta$ factors cannot be achieved in practice. Experimental studies have demonstrated that $\beta > 0.1$ is achievable in single-mode III–V MQW nanowire lasers incorporating multiple quantum wells [15], [16], as well as single-mode quantum-dot micropillar and photonic crystal nanolasers [17], [18]. However, these experimental studies used $\beta$ as a fitting parameter, and the dependence of $\beta$ on parameters such as quantum well thickness, composition and carrier density was not studied.

In this work, we present a model for $\beta$ in MQW nanowire lasers that takes into account the dynamic dependence on charge carrier density, the composition and geometry of the quantum wells, and the properties of the nanowire cavity modes. Our computation combines Finite Element Modeling (FEM) of the electromagnetic field within the nanowire cavity [19] and analysis of the optoelectronic properties of the gain medium starting from the Einstein A and B coefficients. Our results show that $\beta$ is a dynamic quantity that varies with material composition, quantum-well thickness and carrier density.

## II. General Expressions

In this section we derive general expressions for the $\beta$ factor, gain and spontaneous emission rate for cavities containing bulk semiconductors and quantum wells. We consider the most general case, which consists of a nanostructured cavity of refractive index $n_{cav}$ in which are embedded $m$ quantum wells of refractive index $n_{act}$ (see Fig. 1). We assume that the cavity supports a discrete set of $M$ resonant modes at frequencies $\omega_i = E_i/\hbar$ and respective quality factors $Q_i$, with $i = 1...M$.

### A. The $\beta$ factor

The $\beta$ factor is defined as the rate of spontaneous emissions into a particular mode $j$ (usually the lasing mode) divided by the total spontaneous emission rate, $R_{spon}$ [7]:

$$\beta = \frac{R^j_{spon}}{R_{spon}} = \frac{R^j_{spon}}{\sum_{i=1}^M R^i_{spon} + R^{cont}_{spon}} \tag{3}$$

Here, $R_{spon}$ is the rate of spontaneous emissions into the $i^{\text{th}}$ mode in units of $s^{-1}m^{-3}$. The total spontaneous emission $R_{spon}$ in the denominator of Eqn. 3 includes $R^{cont}_{spon}$ which describes the emission rate into the continuum of modes surrounding the cavity.

The spontaneous emission rates are best understood as integrals over the spontaneous emission spectrum $r^j_{spon}(E)$, which gives the emission rate at photon energy $E$ into the $j^{\text{th}}$ cavity mode in units of $J^{-1}s^{-1}m^{-3}$ as:

$$R^j_{spon} = \int_0^\infty r^j_{spon}(E)\mathrm{d}E \tag{4}$$

with a similar definition holding for emission into the continuum $r^{cont}_{spon}(E)$. Starting from the Einstein coefficients for a two level system it can be shown (see chapter 9 of Ref [23]) that the spontaneous emission spectrum can be computed via an integral over the electron energy levels. At thermal equilibrium, the general expression for the spontaneous emission spectrum into the $j^{\text{th}}$ mode is:

$$\begin{aligned} r^j_{spon}(E) &= N^j_{ph}(E) \int_{E_g}^\infty \mathrm{d}E_{cv} C_0(E) \frac{c}{n_{act}} \rho_r(E_{cv}) |\hat{e}\cdot\vec{p}|^2 \times \\ & f_c(E_{cv})\left(1 - f_v(E_{cv})\right)\delta(E_{cv} - E), \end{aligned} \tag{5}$$

Here $C_0(E) = \hbar\pi e^2/n^2_{act}\epsilon_0 m_0^2 E$, with values of the constants given in Table I. To account for linewidth broadening due to intraband scattering, the delta function in Eqn. 5 is replaced by the Lorentzian function with linewidth $\Gamma_{in}$=$\frac{2\hbar}{\tau_{in}}$:

$$Ł(E,\Gamma_{in}) = \frac{\Gamma_{in}/2}{E^2 - (\Gamma_{in}/2)^2} \tag{6}$$

This gives:

$$\begin{aligned} r^j_{spon}(E) &= N^j_{ph}(E) \int_{E_g}^\infty \mathrm{d}E_{cv} C_0(E) \frac{c}{n_{act}} \rho_r(E_{cv}) |\hat{e}\cdot\vec{p}|^2 \times \\ & f_c(E_{cv})\left(1 - f_v(E_{cv})\right) Ł(E - E_{cv}, \Gamma_{in}) \end{aligned} \tag{7}$$

Here $N^j_{ph}(E)$ is the photonic Density of States (PDOS) associated with the cavity mode, $\rho_r(E_{cv})$ is the reduced mass electronic DOS, $|\hat{e}\vec{p}|^2$ is the momentum matrix element, and $f_c(E_{cv})$ and $f_v(E_{cv})$ are the Fermi functions, which depend on the charge carrier density in the active medium via the quasi-fermi levels. The integral extends over all possible values of the transition energy, $E_{cv}$.

From Eqns. (3) and (5), we can write the general expression for the $\beta$ factor as:

$$\begin{aligned} \beta = & \int_0^\infty \mathrm{d}E \int_{E_g}^\infty \mathrm{d}E_{cv} N^j_{ph}(E) C_0(E) \frac{c}{n_{act}} \rho_r(E_{cv}) |\hat{e}\cdot\vec{p}|^2 \times \\ & f_c(E_{cv})\left(1 - f_v(E_{cv})\right) Ł(E - E_{cv}, \Gamma_{in}) \Big/ \\ & \left( \int_0^\infty \mathrm{d}E \int_{E_g}^\infty \mathrm{d}E_{cv} N^{tot}_{ph}(E) C_0(E) \frac{c}{n_{act}} \rho_r(E_{cv}) |\hat{e}\cdot\vec{p}|^2 \times \right. \\ & \left. f_c(E_{cv})\left(1 - f_v(E_{cv})\right) Ł(E - E_{cv}, \Gamma_{in}) \right), \end{aligned} \tag{8}$$

where $N^j_{ph}(E)$ is the photonic density of states (PDOS) nfor the lasing mode, and $N^{tot}_{ph}(E)$ is the total PDOS, given in Eqn. 9).

The existence of resonant cavity modes changes $N^{tot}_{ph}(E)$ at a particular emission energy $E$, from that of free space. The total photonic DOS can be modeled as the free space contribution plus a series of Lorentzians having widths appropriate to each of the cavity modes [13], [24]:

$$N^{tot}_{ph}(E) = \frac{8\pi n^3_{cav} E^2}{h^3 c^3} + \sum_j^M \frac{1}{V_j} L(E - E_j, \Gamma_j) \ , \tag{9}$$

TABLE I
PARAMETERS USED IN THE LASER RATE EQUATIONS.

| Symbol,units | Parameter | Value |
|---|---|---|
| $\eta$ | fraction of pump power that excites carriers [15] | $4.84 \times 10^{-4}$ |
| $f_p, eV$ | energy of pump photons [15] | 1.5498 |
| $V_a$, $m^3$ | volume of active region [15] | $6.59 \times 10^{-21}$ |
| $\beta$ | spontaneous emission factor | |
| $\tau_{sp}$, s | spontaneous emission lifetime | |
| $\Gamma$ | confinement factor [15] | 0.0864 |
| $n$ | refractive index | 3.7 |
| W | number of quantum wells | 10 |
| $L_z$, nm | quantum well width | |
| $\tau_{in}$, s | intraband transition lifetime [20] | $1 \times 10^{-12}$ |
| $E_g(5K)$ | band gap energy of $In_{0.2}Ga_{0.8}As$ at 5K [21] | 1.221 eV |
| $m_e^*$ | effective mass of electron in the conduction band [22] | $0.0582m_0$ |
| $m_h^*$ | effective mass of hole in the valence band [22] | $0.500m_0$ |
| $d$, nm | nanowire diameter [15] | 200 |
| $L$, nm | length of the nanowire along optical axis [15] | 2200 |

where $\Gamma_j = E_j/Q_j$, and $L$ is the Lorentzian function.

The PDOS is the number of photonic states available at a specific transition energy per unit volume and is dependent on the cavity structure (geometry, size, composition). Through the PDOS, the model incorporates the influence of the cavity on the electronic transitions between the conduction and valence bands and subsequent coupling of the resulting photonic emission into the lasing mode. The model (Eqn. 9) for the PDOS involves the replacement of an infinite set of modes in the continuum, including all cavity modes, with a free-space term (the first term in (Eqn. 9) plus the additional set of Lorentzian cavity modes which are considered separate from the free-space part. This model will apply in situations where the cavity modes are sufficiently well-confined to the high-index region that they can be considered to be cleanly separated from the free-space continuum, and where the cavity modes form a discrete set with a discernible frequency separation between them. The calculated PDOS for the MQW nanowire structure of Fig. 1 is shown in Fig. 2.

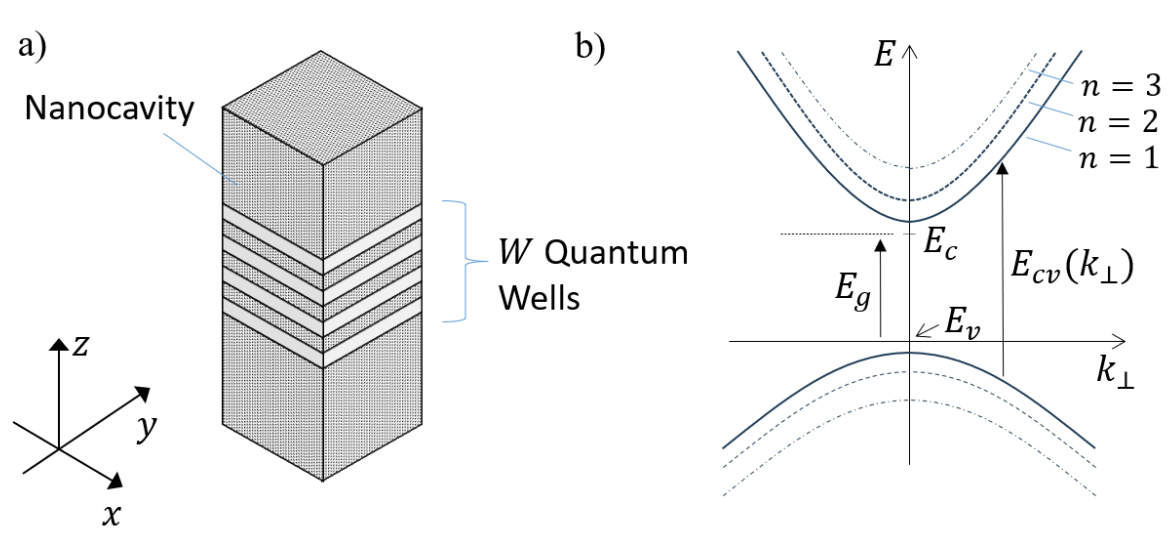

Fig. 1. (a) Schematic of multiple quantum well semiconductor nanowire laser. We assume that the quantum wells are oriented perpendicular to the $z$ axis and have equal widths $L_z$; b) Energy level diagram showing the first three energy levels in the quantum wells as a function of the perpendicular wavenumber $k_\perp$. The edge of the valence band in the bulk active semiconductor is set at $E_v = 0$ throughout this paper. The band edges shift due to quantization introduced by the quantum well.

The plane of the quantum wells in Fig. 1 are aligned perpendicularly to the $z$-axis and have thickness ($L_z$) and composition $In_{0.2}Ga_{0.8}As$). The temperature is constant at 5 K, and the effective electron and hole masses are $m_e^*$ and $m_h^*$, with values given in Table I. The band diagrams are shown in schematic form in Fig. 1b. The wells have an electronic density of states given by:

$$\rho_{2D}(E') = \sum_q H(E - E_g - E_q)\frac{m_z}{\pi\hbar^2 L_z} \quad (10)$$

where $E_q = \frac{\hbar^2}{2m_e}(\frac{q\pi}{L_z})^2$.

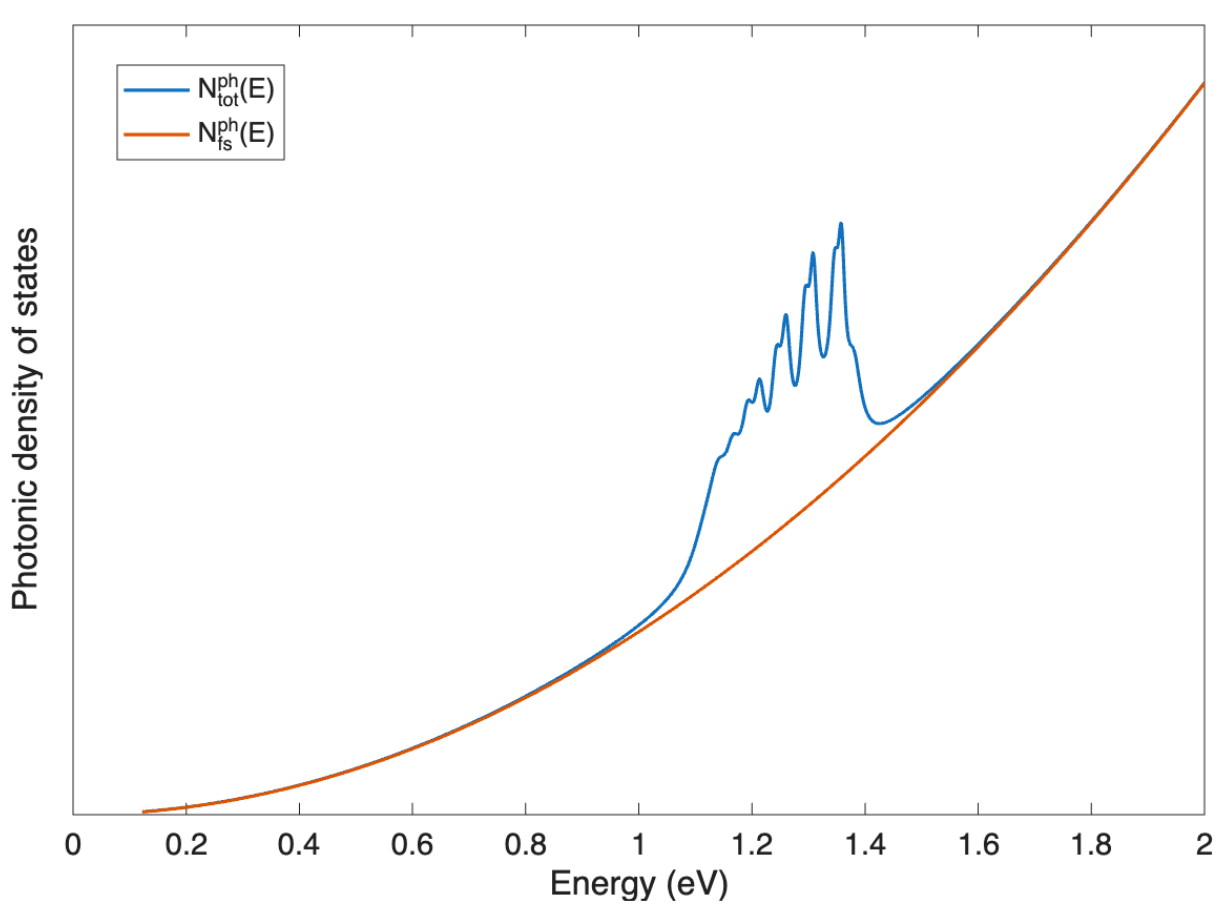

Fig. 2. Total photonic density of states $N_{tot}^{ph}(E)$ of the MQW nanowire laser depicted in Fig. 1. The free space density of states $N_{fs}^{ph}(E)$ is also shown.

## B. Gain

It can be shown [23] that the gain at an emission energy $E$ for a single quantum well in such a nanowire structure is given by:

$$\begin{aligned} g(E, N_c) &= C_0(E)\int_{E_g}^{\infty} \mathrm{d}E_{cv}\rho_r(E_{cv})|\hat{e}\cdot\vec{p}|^2 \\ &\times (f_c(E_{cv}, N_c) - f_v(E_{cv}, N_c))\, L(E, E_{cv}) \end{aligned} \quad (11)$$

in units of $m^{-1}$. Here $C_0(E)$ is a constant, $\rho_r(E_{cv})$ is the reduced electronic density of states, $|\hat{e}\cdot\vec{p}|^2$ is the momentum matrix element, $f_c(E_{cv}, N_c)$ and $f_v(E_{cv}, N_c)$ are the Fermi functions, and $L(E, E_{cv})$ is the Lorentzian function. To arrive

at this expression, several assumptions have been made: first, it is assumed that the transition processes are dominated by heavy-hole interactions. This is justified by empirical data [25].

Second, it has been assumed that transitions only occur between electron and hole states with equal quantum numbers. This involves the insertion of an overlap integral [23], in which the factors are exactly zero for the first two transitions and become very small thereafter. We note that the gain does not depend on the photonic DOS. It does, however, depend strongly on the carrier densities via the Fermi functions $f_c(E_{cv})$ and $f_v(E_{cv})$.

### C. $\beta$ factor and gain for a quantum well

The equations derived thus far apply to bulk semiconductors. Without any loss in generality, they can be used for quantum wells by replacing the 3D electronic density of states function $\rho_r(E_{cv})$ in Eqn. 8 with a 2D electronic density of states function and integrating over the QW volume in $k$ space, as given in Eqn. 12:

$$\frac{1}{L_z(2\pi)^2}\sum_{\vec{k_a}}\Delta k_t \rightarrow \frac{1}{L_z(2\pi)^2}\iint d\vec{k_t} \qquad (12)$$

Quantization of electron and hole energy levels within the conduction and valence bands due to 1D confinement in the quantum wells leads to a step-like density of states function and pushes the first available conduction and valence band states away from the band edge, effectively increasing the band gap (Fig. 3). The separation between the quantized electron and hole energy states is dependent on the quantum well width, $L_z$.

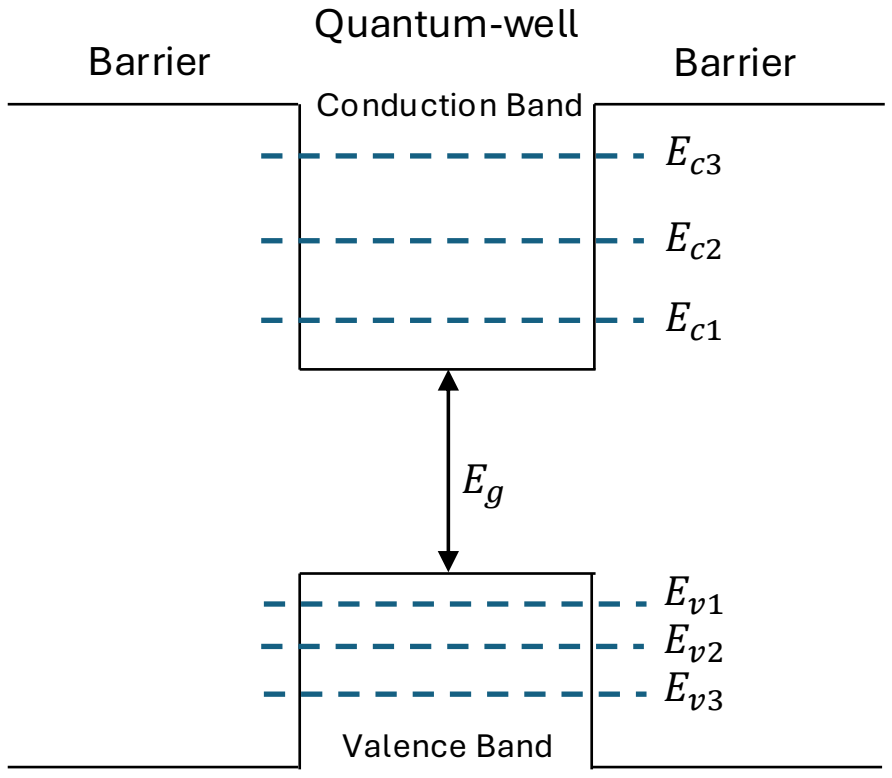

Fig. 3. Three discrete electron $(E_{c1}, E_{c2}, E_{c3})$ and hole $(E_{v1}, E_{v2}, E_{v3})$ energy levels within the conduction and valence bands of each QW.

## III. Methods

### A. Cavity Simulations

Fig. 4 illustrates the hexagonal nanowire laser cavity, which has length of 2.2 $\mu$ m and diameter 200 nm, and is composed of ten $In_{0.2}Ga_{0.80}As/GaAs$ quantum wells, each with thickness 19 nm. The structure is based on the single mode nanowire laser reported in Ref. [15], which was fabricated by selective area epitaxy using a hexagonal hole array with a diameter of 200 nm. The nanowire cavity is optically pumped from above by an 800 nm pulsed laser with pulse width of 35 fs. We assume that the pump pulse has a Gaussian shape $P(t) = P_0 exp(\frac{-t^2}{\Delta t^2})$, where $P_0$ is the peak power and $\Delta t$ is the pulse width.

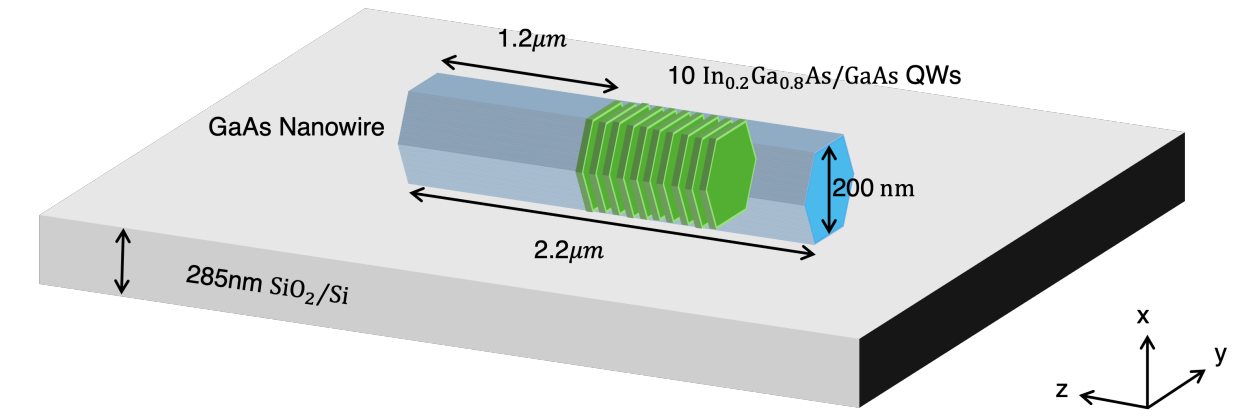

Fig. 4. Schematic diagram of the modelled $In_{0.2}Ga_{0.80}As/GaAs$ quantum-well nanowire laser. The nanowire cavity is surrounded by air and contains $W$=10 quantum wells, each with thickness 19 nm.

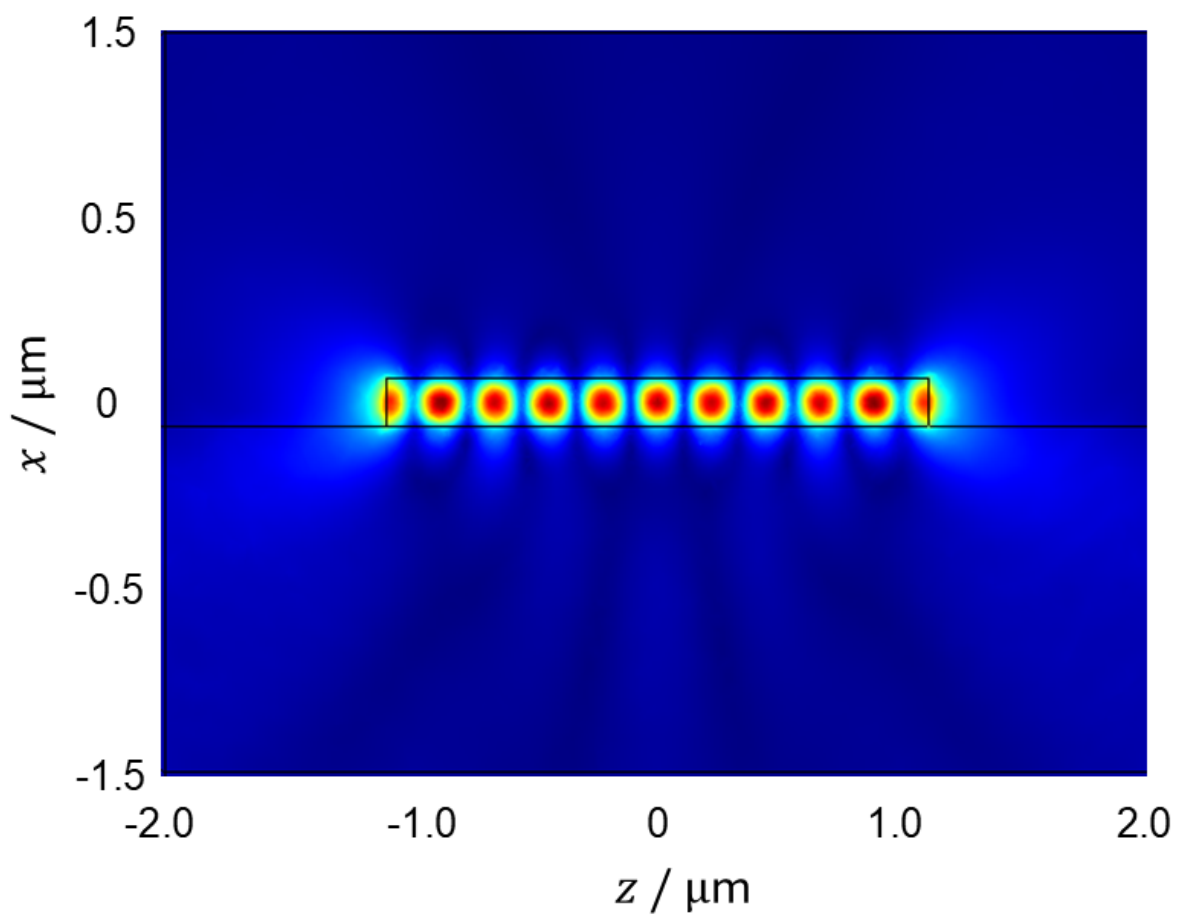

Fig. 5. Cross-section from the 3D cavity simulation of the $In_{0.2}Ga_{0.80}As/GaAs$ MQW nanowire laser showing the electric field intensity of the HE11b lasing mode along the longitudinal (x) axis.

Cavity simulations of the MQW nanowire were conducted using the COMSOL Multiphysics package using the finite element method, enabling identification of the longitudinal and transverse modes of the nanowire laser. The longitudinal modes determine the frequencies at which the laser can oscillate, while transverse modes describe the intensity distribution of the laser beam's cross-section. The COMSOL simulations of the resonant modes and frequencies of this nanowire cavity reveal that only two lasing modes, HE11a and HE11b, are supported in the cavity, confirming the results of the Lumerical simulations in Ref. [15]. However, only the HE11b mode is polarized in the transverse direction to the nanowire axis, as observed experimentally (Fig. 2(b) of ref. [15]). This mode has a resonant frequency of 312.85 GHz. Its' field profile as obtained from the 3D COMSOL simulation is shown in Fig. 5.

## IV. Effect of dynamics and quantization on lasing behaviour

We present a validation of our model for the MQW nanowire cavity of Fig. 4 in Appendix 3. We demonstrate consistency with experimental results for low-temperature operation of a MQW, high-index nanowire [15]. Although the model can be expected to apply at higher temperatures and to a broader family of quantum well lasers, the formal validation of the model for these structures would require additional fabrication and experiments. We leave such a comparison to future work.

In this section, we present results that extend the experimental measurements made in Zhang et al. [15] to show how our model can be used to predict laser performance variation with charge carrier density, quantum well thickness and material composition. The results discussed in this and the following sections are for $W$=10 quantum wells placed in a nanowire cavity that is surrounded by air (based on the nanowire laser studied by Zhang et al. [15]). Nanowire dimensions, material compositions and other experimental parameters are given in Table I with references where appropriate. The well composition is $In_{0.2}Ga_{0.8}As$, the barriers are GaAs, the refractive index $n = 3.7$ for both wells and barrier, and the quantum well width $L_z$ is varied. We assume that the temperature is 5 K as in the experiment, and that the nonradiative lifetime is much greater than the radiative lifetime (which is justified by the surface passivation provided by encapsulation within the AlGaAs/GaAs shell). Thus the effects of varying temperature and nonradiative recombination are not included in the model at present.

### A. Dynamic nature of $\beta$ and dependence on charge carrier density

The expressions for $\beta$ and gain in Sec. II depend on charge carrier density through the Fermi functions which are given by:

$$f_c(E_{cv}) = \frac{1}{1+exp[(E_g+\frac{m_v^*}{m_c^*}(E_{cv}-E_g)-F_c)/k_bT]} \quad (13)$$

$$f_v(E_{cv}) = \frac{1}{1+exp[-\frac{m_v^*}{m_c^*}(E_{cv}-E_g)}) - F_v] \quad (14)$$

where $E_{cv}$ is the transition energy between the valence and conduction bands and $F_c$ and $F_v$ are the Quasi-Fermi energies of the conduction and valence band. These energies change with the carrier population as the laser is operated; for a volume density $N_c$ of populated valence electrons (with an equal volume density $N_v$ of holes), the quantities $F_c$ and $F_v$ can be computed via equations 15 and 16:

$$N_c = \int_{E_g}^{\infty} dE' \rho_c(E_0) f_c(E_0) \quad (15)$$

and

$$N_v = \int_{-\infty}^{E_v} dE' \rho_v(E_0) f_v(E_0) \quad (16)$$

The dependence of the Fermi functions $f_c(E_{cv})$ and $f_v(E_{cv})$ on carrier density directly reflects the position of the electron and hole quasi-Fermi levels, $F_c$ and $F_v$. At low carrier densities, where there is no excitation, $F_c$ lies below the conduction-band edge and $F_v$ lies above the valence-band edge, so $f_c(E_{cv}) \ll 1$ and $f_v(E_{cv}) \approx 1$ near the band edge, and the term $f_c(E_{cv})\big[1 - f_v(E_{cv})\big]$ remains close to zero, thus there is no population inversion. As the carrier density increases, $F_c$ shifts upward into the conduction band and $F_v$ shifts downward into the valence band, leading to $f_c(E_{cv})$ approaching unity and $f_v(E_{cv})$ decreasing below unity over a finite energy range. In this regime $f_c(E_{cv})\big[1 - f_v(E_{cv})\big]$ becomes large, which corresponds to the build-up of population inversion and the emergence of positive optical gain. At high carrier densities, the increasing separation $F_c - F_v$ broadens the energy range where $f_c(E_{cv}) > f_v(E_{cv})$, while carrier-induced band filling (Pauli blocking) drives $f_c(E_{cv}) \to 1$ and $f_v(E_{cv}) \to 0$ near the band edge, so $f_c(E_{cv})\big[1 - f_v(E_{cv})\big]$ tends to saturate, reflecting the saturation of the achievable gain. Thus, the dynamic nature of $\beta$ can be appropriately modeled by the inclusion of charge carrier effects as described by Eqn. 8.

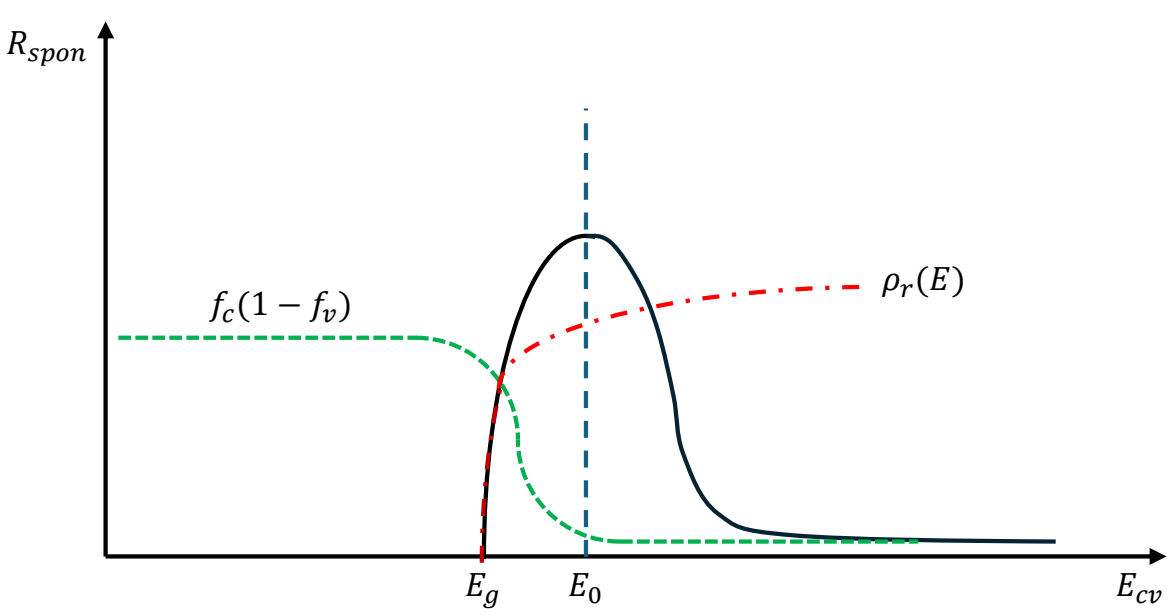

Fig. 6. Relationship between different terms in the total spontaneous emission rate $R_{spon}$ ($s^{-1}m^{-3}$). The black line is $R_{spon}$, the occupancy probability $f_c(1-f_v)$ is shown with the green dashed line and the 2D electronic density of states function $\rho_r(E)$ is shown with the red dot-dashed line.

For the laser structure of Fig. 1, the spontaneous emission factor $\beta$ (Eqn. 3) is obtained by dividing the area under the spontaneous emission spectrum of the lasing mode, $R^j_{\text{spon}}$ by the sum of the areas under the spontaneous emission spectrum of the nanowire cavity $R_{\text{sp,cav}}$ and of the continuum $R^{cont}_{spon}$. The dependence of this dynamic spontaneous emission factor $\beta$ on carrier density $N_c$ is plotted as the blue curve in Fig. 7 and the 'static' value of $\beta$=0.111 calculated from Eqn. 2 is plotted as the red line. At carrier densities of $N_c = 1 \times 10^{23} m^{-3}$ and below (red line in Fig. 7), where $N_c$ is close to the intrinsic carrier density, the denominator of Eqn. 3 is much larger than the numerator. Thus, $\beta$ is small ($\beta \approx 0.02$) (Fig. 7). As $N_c$ increases to $10^{24}$ (green curve in Fig. 8), $\beta$ increases and reaches a maximum value of 0.110 at the threshold carrier density $N_c = 7.92 \times 10^{23} m^{-3}$ (Fig. 7). In this regime, a relatively large fraction of the total spontaneous emission is channeled into the lasing mode. In the following section we elucidate the dependence of $\beta$ on carrier density and quantum well thickness.

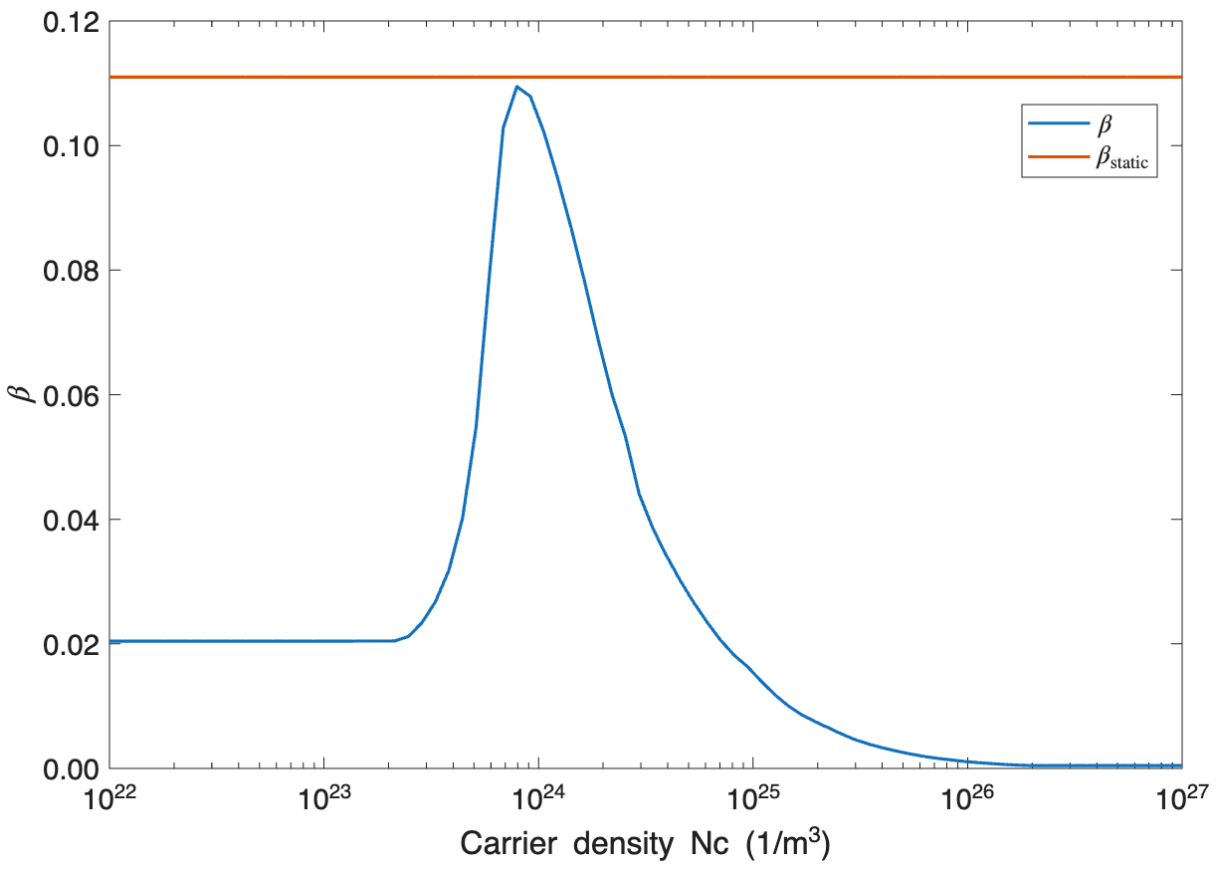

Fig. 7. Dependence of $\beta$ on charge carrier density $N_c$ for the multiple quantum well nanowire laser depicted in Figure 1. The quantum wells have widths of $L_z$= 19 nm and composition $In_{0.2}Ga_{0.8}As$) as per Ref. [15].

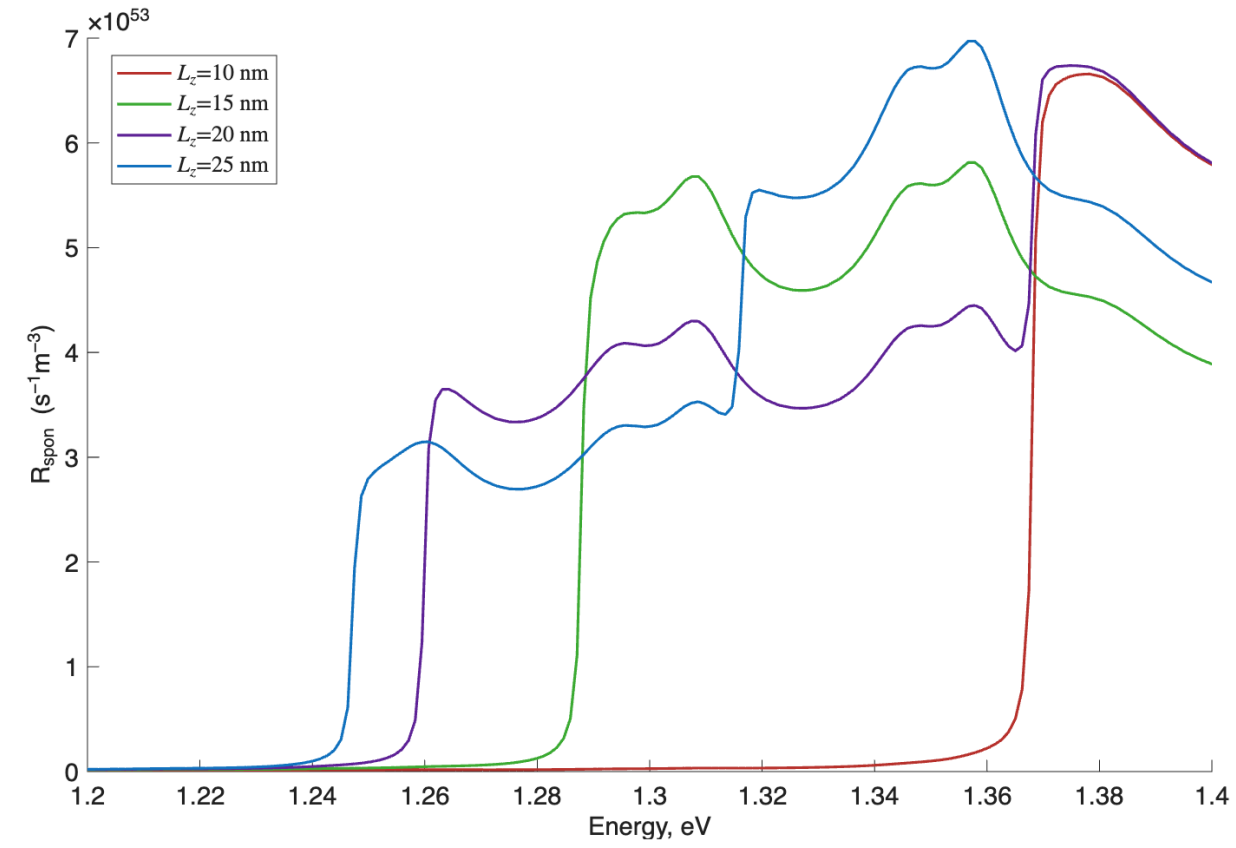

Fig. 9. Spontaneous emission spectrum ($R_{spon}$) for quantum well widths $L_z$ equal to 10-25 nm at a carrier density of $N_c = 10^{27} m^{-3}$.

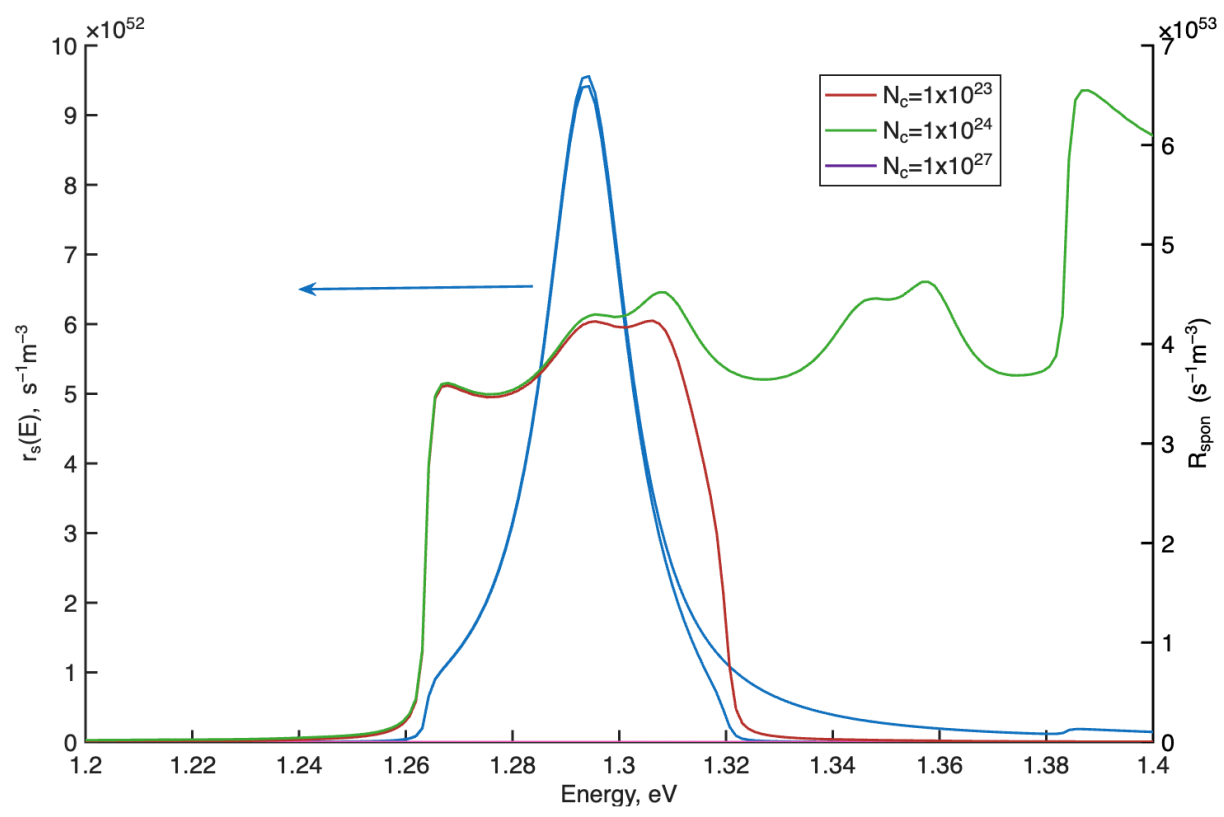

Fig. 8. Spontaneous emission spectrum from the lasing mode $r^j_{spon}(E)$ (left y axis) and total spontaneous emission spectrum ($R_{spon}$) (right y axis) for carrier densities $10^{23} - 10^{27} m^{-3}$ and a quantum well width of $L_z$=19 nm as in Ref. [15].

### *B. Impact of quantization on spontaneous emission and lasing characteristics.*

Fig. 8 shows the spontaneous emission spectrum from the lasing mode $r^j_{spon}(E)$ for carrier densities ranging from $10^{23} m^{-3}$ (bottom) to $10^{27} m^{-3}$ on the left-hand y axis, with the total spontaneous emission rate $R_{spon}$ ($s^{-1} m^{-3}$) on the right-hand y axis. The total spontaneous emission spectrum shows that the first and second energy levels in the quantum wells are populated at carrier densities equal to or greater than $10^{23} m^{-3}$ (red curve). The third and fourth energy levels become populated as the carrier density increases to $10^{24} m^{-3}$ and $10^{27} m^{-3}$ (overlapping green and purple curves). As the carrier density increases from $10^{24} m^{-3}$ to $10^{27} m^{-3}$, the change in $r^j_{spon}(E)$ is very small compared to the change in the free space and total nanowire spontaneous emission spectrum $R_{spon}$. Thus for higher charge carrier densities the spontaneous emission is more likely to couple to other modes of the nanowire cavity or to the free space mode continuum than to the lasing mode, and the contribution of the denominator of Eqn. 3 increases substantially more than the numerator, leading to the decrease in $\beta$ above threshold observed in Fig. 7.

The contributions of higher energy levels to $R_{spon}$ at a carrier density of $10^{27} m^{-3}$ can also be seen with increasing quantum well width $L_z$ from 10 to 25 nm in Fig. 9. The variation of $\beta$ with carrier density is shown in Fig. 10 for quantum well widths ranging from 10 to 25 nm. For all quantum well widths, $\beta$ peaks just above the threshold carrier density $N = 7.92 \times 10^{23}\ m^{-3}$. Notably, there is an optimum quantum well width (in this case 15 nm, plotted with the green dashed line) for which $\beta$ is maximized at a value of 0.19, almost double the 'static' value of 0.111 calculated from Eqn. 3. In contrast, the optimal value of $\beta$ does not vary greatly with QW composition ranging from 10-30% Indium, as shown in Fig. 11. Future development of our model will incorporate temperature dependence of the bandgap as well as nonradiative recombination. This will allow us to determine the optimal quantum well widths for epitaxial semiconductor heterostructures with different quantum well and barrier compositions.

Fig. 12 shows that for the MQW nanowire laser illustrated in Fig. 4, there is a larger jump in the emitted pulse energy below and above the threshold when $\beta$ is dynamic (i.e., for $\beta(N)$) compared to constant $\beta$). This occurs because the value of $\beta$ lies below the static value of $\beta$ for most values of the carrier density (see Fig. 10), and hence for the majority of pump fluences in Fig. 12. In addition, the threshold shifts to slightly higher pump fluence. The larger jump in the dynamic $\beta$ case also results in a steeper slope at threshold. These changes occur because there are fewer excited states emitting into the lasing mode below the threshold power for $\beta(N)$ compared to constant $\beta$.

## V. Conclusion

In this paper we have presented a model for computation of dynamic lasing performance starting with Einstein's coefficients, and calculate the spontaneous emission rate, the $\beta$ factor, and gain for a semiconductor multiple quantum well laser. Previous work in this area has generally assumed $\beta$ to be constant; here we show strongly varying behavior in the vicinity of threshold. In contrast to those authors who have acknowledged $\beta$ as a dynamic quantity (notably ref. [8]) our model makes no underlying assumptions about the rate of spontaneous emission into continuum or leaky modes. The model approximates the full continuum of electromagnetic modes about the cavity by a direct sum of leaky cavity modes with modes from free space, and is therefore limited to situations where the cavity modes are relatively distinct and well-confined to the high-index region. As an example of the utility of our model for improving laser design, we optimize the carrier density, composition and quantum well width in a typical MQW-nanowire laser so as to maximize the spontaneous emission factor $\beta$.

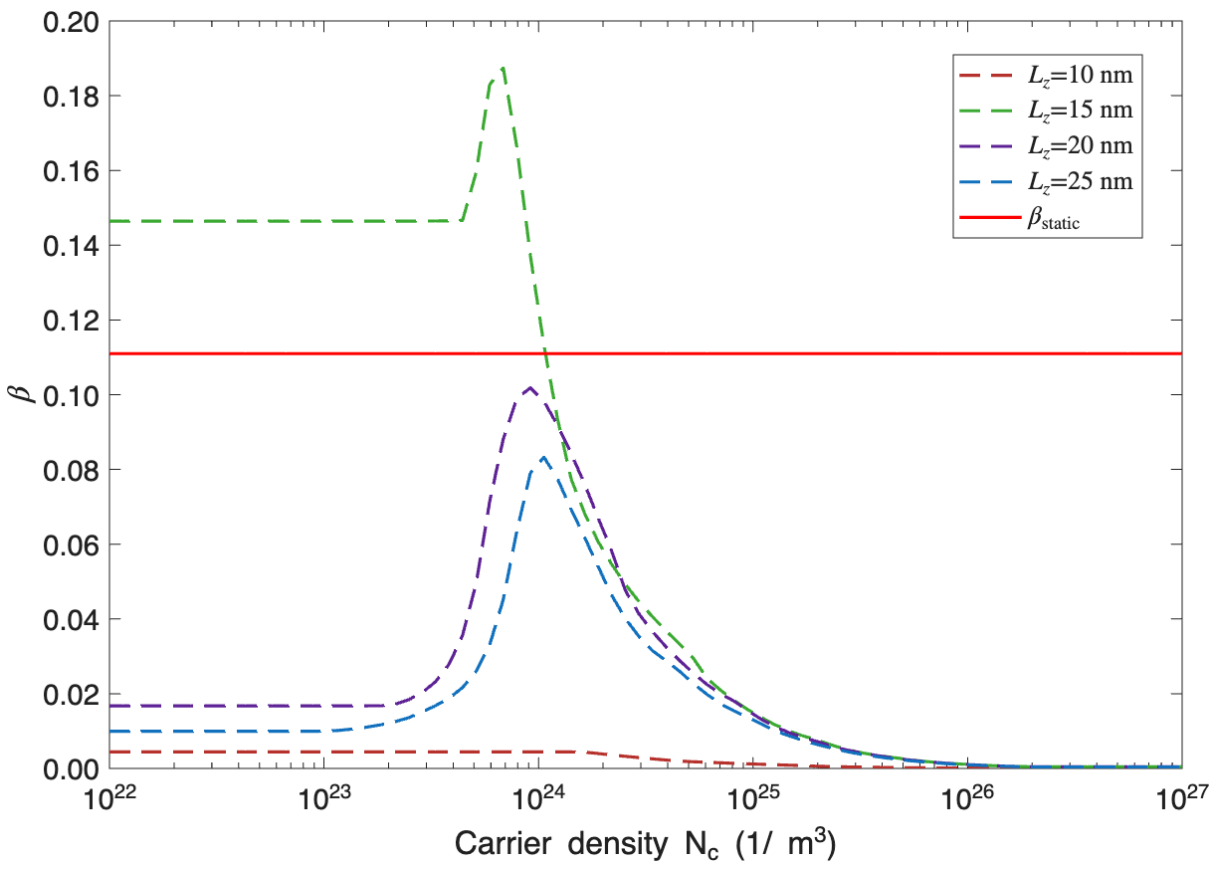

Fig. 10. Dependence of $\beta$ on quantum well width and charge carrier density $N_c$ for the multiple quantum well nanowire laser depicted in Figure 1. The quantum wells have widths varying from $L_z$= 10-25 nm.

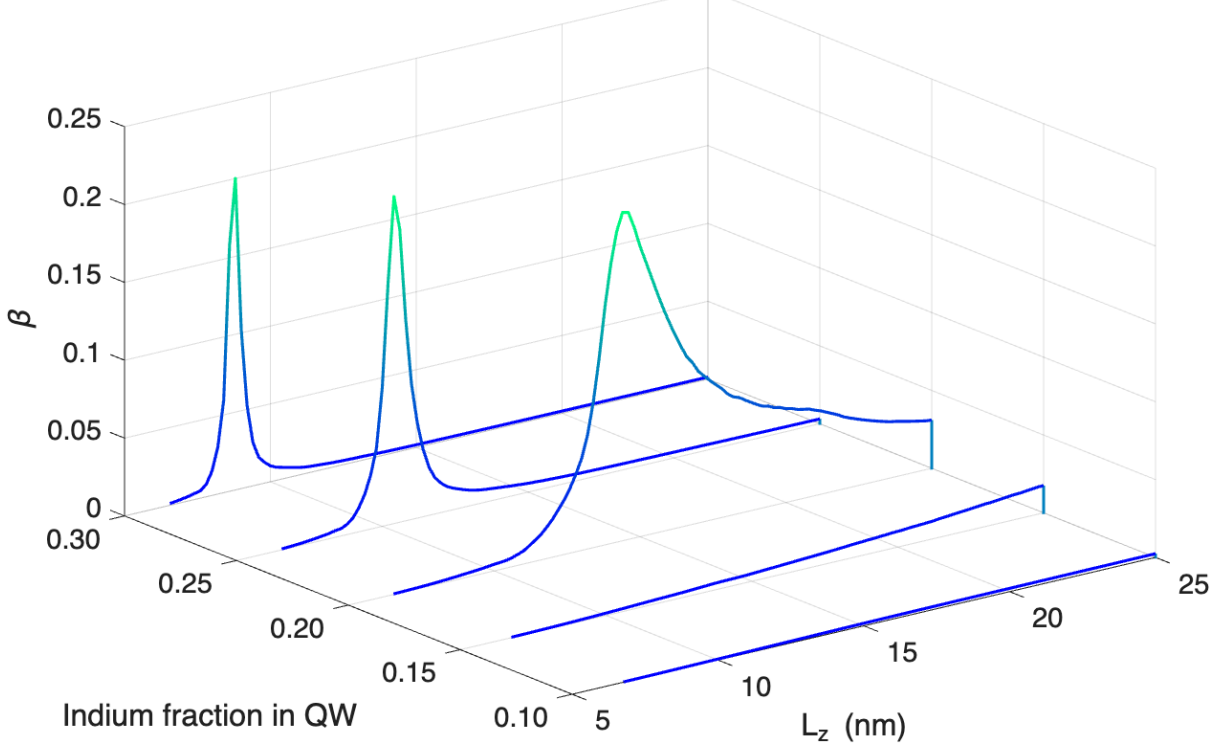

Fig. 11. Dependence of $\beta$ on Indium fraction (0.10-0.30) and thickness ($L_z$=5-25 nm) of the quantum wells.

## Appendix 1. Computation of $\beta$, gain and spontaneous emission rate for bulk and quantum wells

In this section we present some of the key steps in the derivation for the $\beta$ factor without going into every detail, and provide the equations for computation of the gain and spontaneous emission rate as well.

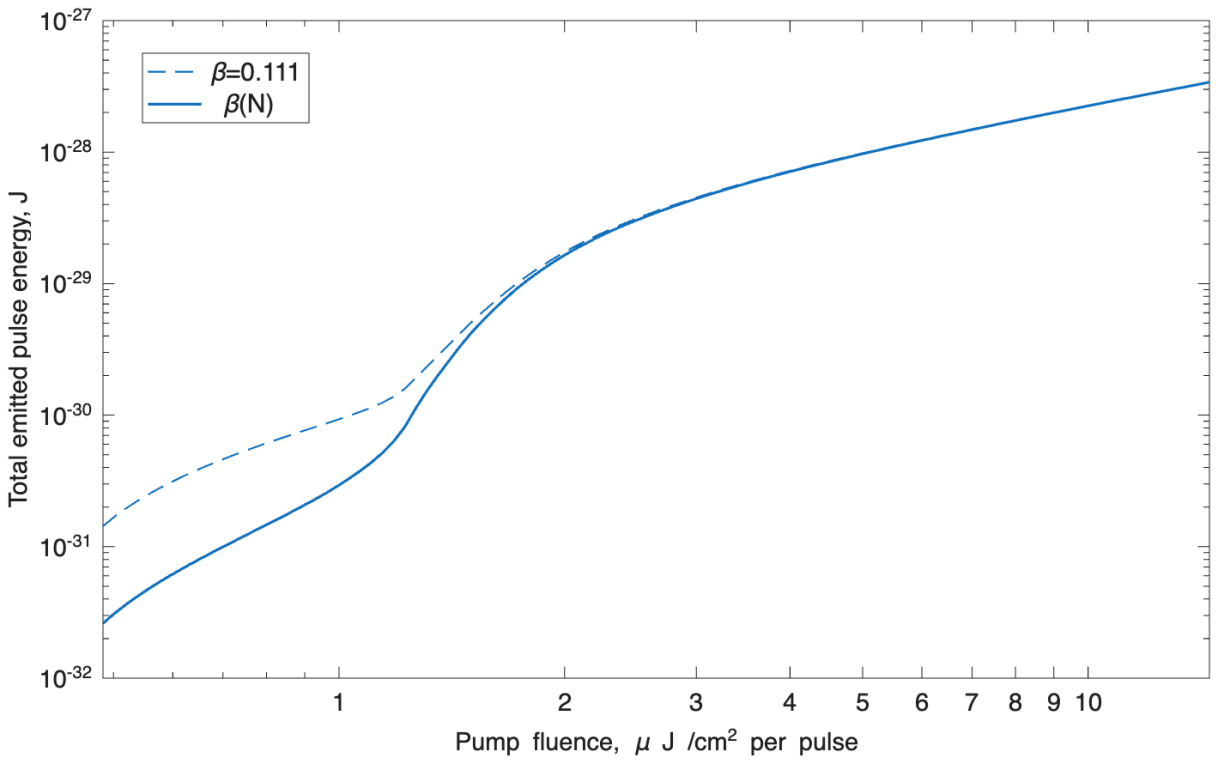

Fig. 12. Light in-light out curve at 5 K for the MQW nanowire laser illustrated in Fig. 4. For both static and dynamic $\beta$, the laser threshold occurs at a pump power of approximately 1.5 $\mu J/cm^{-2}$ per pulse, in agreement with the experimental laser threshold of 1.6 $\mu J/cm^{-2}$ per pulse (Fig. 2c of ref. [15]).

### A. $\beta$ factor in a bulk semiconductor

To calculate the $\beta$ factor and gain, we require the spontaneous emission rates into the lasing mode $R^{j}_{spon}$ and into the continuum, $R^{cont}_{spon}$. These were defined in Eq. 5.

Following [23] and starting with Einstein's A and B coefficients for a discrete two-level system we have:

$$R^{abs}_{12} = R^{stim}_{21} + R^{spon}_{21} \tag{17}$$

which leads to

$$B_{12}P(E_{12})f_1(1-f_2) = B_{21}P(E_{21})f_2(1-f_1)+A_{21}f_2(1-f_1) \tag{18}$$

where $A$, $B$ are the Einstein's A and B coefficients, $f_1$ and $f_2$ are the Fermi functions, and $P(E)$ is the energy density of photons.

Thus we arrive at the well known result:

$$P(E_{12}) = \frac{A_{21}}{B_{12}e^{E_{21}/k_BT} - B_{21}} \tag{19}$$

For Eq. 19 to be hold true at all temperatures $B_{12} = B_{21}$ and $\delta F = 0$ which gives us:

$$\frac{A_{21}}{B_{21}} = N^{tot}_{ph}(E_{21}) \tag{20}$$

These results are for discrete energy states. For lasing in semiconductor materials such as the one considered here, we also have to account for electronic transitions involving the populated continuum of states (as shown in Fig. 1).

This gives us:

$$R^{absnet}_{12} = \int_0^\infty B_{21}N^{tot}_{ph}(E_{21})(f_1-f_2)\delta(E_2-E_1-E)N_{ph}dE, \tag{21}$$

where $f_1$ and $f_2$ are the Fermi functions defined as

$$f_1(E_1) = \frac{1}{1+e^{(E_1-E_f)/k_BT}} \tag{22}$$

$$f_2(E_2) = \frac{1}{1+e^{(E_2-E_f)/k_BT}} \tag{23}$$

Summing over all the electronic states with wavevectors $\vec{k_a}$, $\vec{k_b}$ and accounting for spin degeneracy of the electrons we obtain:

$$r_{net}^{abs}(E) = 2N_{ph}^{tot}(E)\sum_{\vec{k_a}}\sum_{\vec{k_b}} B_{ba}(f_b - f_a)\delta(E_b - E_a - E) \tag{24}$$

We use the expression for $B_{ab}$ from [23]:

$$B_{ab} = \frac{\pi e^2}{n_r^2\epsilon_0 m_0^2\omega}|\hat{e}\cdot p_{ba}|^2\frac{1}{V} \tag{25}$$

and $\{A^2 = \frac{2\hbar}{n_r^2\epsilon_0\omega}\frac{1}{V}$, where $A_0$ is the electromagnetic vector potential arising from a single photon in a volume $V_1$ and which we can think of as a volume contained within the active region to obtain:

$$B_{ba} = C_0(\frac{c}{n})(\frac{1}{V})|\hat{e}\cdot \vec{p_{ba}}|^2\delta_{\vec{k_a},\vec{k_b}} \tag{26}$$

where $C_0$ is

$$C_0 = \frac{\hbar\pi e^2}{n_r c\epsilon_0 m_0^2 E} \tag{27}$$

This results in the following expression for the spontaneous emission rate:

$$r^{spon} = 2N_{ph}^{tot}(E)C_0(E)(\frac{c}{n})(\frac{1}{V})\sum_{\vec{k_a}}\sum_{\vec{k_b}}|\hat{e}\cdot \vec{p_{ab}}|^2\delta_{\vec{k_a},\vec{k_b}} \times f_b(1-f_a)\delta(E_b - E_a - E) \tag{28}$$

Conservation of momentum implies $\vec{k_a} = \vec{k_b}$. Then we can write $E_{cv}(\vec{k})$ as

$$E_{cv}(\vec{k_a}) = (E_a - E_b)|_{\vec{k_a}=\vec{k_b}} = E_c(\vec{k_a}) - E_v(\vec{k_b}) \tag{29}$$

which restricts us to those electron-hole pairs between which a direct transition is possible. For a 3D active region, the sum in Eq. 28 can be replaced by an integral as

$$\frac{1}{(2\pi)^3}\sum_{\vec{k_a}}\Delta k = \frac{1}{(2\pi)^3}\iiint d\vec{k} \tag{30}$$

To move from $\vec{k}$ space we need to change variables in the equation above to $E$. In 3D we have:

$$\int d\vec{k}.g(k) = (4\pi)\int_0^\infty k^2 g(k)dk \tag{31}$$

$$= 4\pi^3\int_{E_g}^\infty dE_{cv}\rho_r(E_{cv})g(k(E_{cv})) \tag{32}$$

This brings us to the final results of Eq. 8 for a bulk semiconductor active region in a cavity.

### *B. Gain in a bulk semiconductor*

The gain is related to the net absorption in the active material and can be expressed as:

$$R_{12}^{abs-net} = R_{12}^{abs} - R_{21}^{stim} \tag{33}$$

and with the following steps similar to the derivation of $\beta$ we obtain

$$r^{net-abs}(E) = 2\sum_{\vec{k_a}}\sum_{\vec{k_b}} B_{ab}N_{ph}^{tot}(E)(f_a - f_b). \delta(E_b - E_a - E) \tag{34}$$

.

Because the absorption is independent of the electron and hole wave vectors we can write this as:

$$r^{net-abs}(E) = 2N_{ph}^{tot}(E)\sum_{\vec{k_a}}\sum_{\vec{k_b}} B_{ab}(f_a - f_b)\delta(E_b - E_a - E). \tag{35}$$

The absorption spectrum within a spectral width $\Delta E$ is thus:

$$\alpha\cdot\Delta E = \frac{\text{no. photons absorbed/volume/s}}{\text{no. photons injected/area/s}} \tag{36}$$

which can be written as

$$\alpha(E) = 2(\frac{n}{c})\sum_{\vec{k_a}}\sum_{\vec{k_b}} B_{ab}(f_a - f_b)\delta(E_b - E_a - E). \tag{37}$$

In the expression for $\alpha$ we replace $B_{ab}$ and for 3D space replace the summation in $k$ space to obtain:

$$\alpha(E) = C_0\int_{E_g}^\infty dE_{cv}\rho_r(E_{cv})|\hat{e}\cdot \vec{p_{cv}}|^2(f_v - f_c)L(E, E_{cv}). \tag{38}$$

The gain, $g$, is simply $-\alpha$.

## Appendix 2. Laser rate equations

We build on the laser rate equations previously developed for a single-mode nanoscale semiconductor laser under electrical pumping [26], replacing the electrical pump $I$ with an optical pump $P(t)$ with energy $f_p$ eV. The laser rate equations consist of two coupled equations for the rate of change of the number of carriers $N$ in the cavity, and the rate of change of the number of photons $S$. In order to model the effect of carrier density $N$ on the lasing threshold and gain g(N), we include the spontaneous emission factor $\beta$(N) in the second term of equations 39 and 40. Values of constants used in these equations are given in Table I.

$$\frac{dN}{dt} = \frac{\eta P(t)}{hf_p}.\frac{1}{V_a} - \frac{n}{\tau_{sp}} - \Gamma g(N).\frac{S}{V_a} \tag{39}$$

$$\frac{dS}{dt} = -\frac{S}{\tau_p} + \beta(N)\frac{nV_a}{\tau_{sp}} + \Gamma.g(N).S \tag{40}$$

## Appendix 3. Model validation

In this section we present in brief a validation of our model by calculating measurable output quantities from the model (laser threshold, L-L curves, $R_{sp}$) and comparing these with the published experimental results in [15]. The nanowire cavity simulation has been described in Sec. II, Section A which shows excellent agreement with results in ref. [15].

Further, after obtaining the lasing mode we solve the laser rate equations given in Appendix 2 to yield the spontaneous emission rate of the lasing mode $R_{sp}$. This is shown in Fig. 13 for carrier densities between $1\times10^{23}$ and $1\times10^{27}m^{-3}$. This shows that carrier densities equal to or greater than the threshold carrier density $N_c$ = 7.92x $10^{23}m^{-3}$ are required to achieve lasing from the device structure shown in Fig. 1. With increasing $N_c$ above threshold up to $10^{27}m^{-3}$, the spectrum broadens and develops a low wavelength tail. These changes are due to band filling and the separation of the electron and hole quasi-Fermi levels, which move deeper into the conduction and valence bands, introducing additional radiative transitions at higher photon energies.

In ref. [15] the emission spectrum was reported to extend from 950 to 1000 nm (Fig. 2b of ref. [15] and the lasing wavelength was reported to be centred at 959 nm. Comparing Fig. 13 with Fig. 2b of Ref. [15] it can be seen that nanowire emission spectrum calculated with our model is centred at wavelength $\Delta\lambda_{em} = 958$ nm, in agreement with the experimental results.

Further, we calculate the laser threshold which can be obtained from the light in-light out (L-L) curve, which plots the lasing emission intensity versus the pump power ($\mu J/cm^2$ per pulse). The L-L curve plotted in Fig. 12 shows that for both constant (static) and dynamic $\beta$, the laser threshold is predicted to occur at a pump power of approximately 1.5 $\mu J/cm^{-2}$ per pulse, which is in excellent agreement with the experimental laser threshold of 1.6 $\mu J/cm^{-2}$ per pulse (Fig. 2c of ref. [15].).

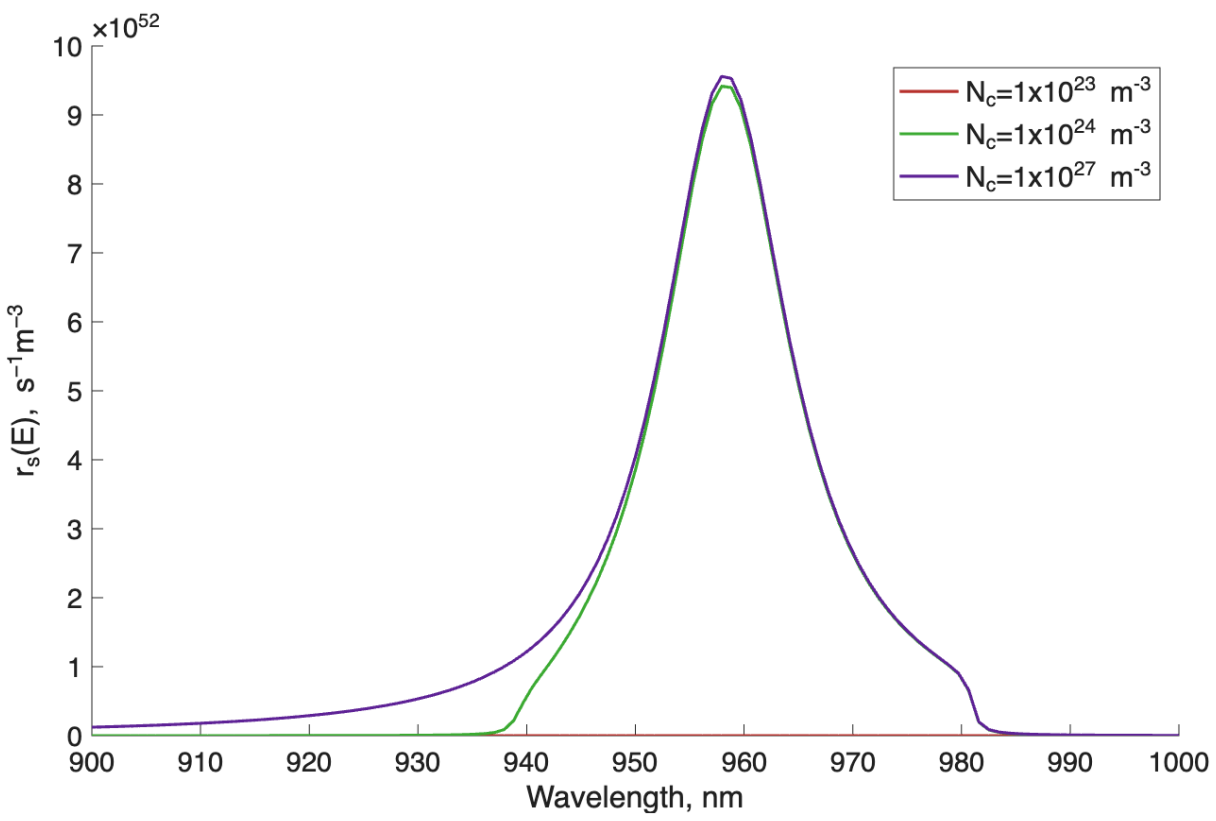

Fig. 13. Spontaneous emission spectrum $r_{sp}(E)$ of the lasing mode at 5K for carrier densities ranging from $10^{23}m^{-3}$ to $10^{27}m^{-3}$ for the MQW nanowire laser illustrated in Fig. 4. Lasing is observed for $N_c > 10^{23}m^{-3}$. The emission peaks at 958 nm, in agreement with the experimental emission spectrum reported in Fig. 2b of Ref. [15].